\documentclass[aps, prd, preprint, nofootinbib, amsfonts, floatfix]{revtex4}

\usepackage{graphicx}
\usepackage{subcaption}
\usepackage{amsmath,amsfonts}
\usepackage{amssymb}
\usepackage{dsfont}
\usepackage{url}
\usepackage{braket}
\usepackage{slashed}
\usepackage{epsfig,color}
\usepackage[dvipsnames]{xcolor}
\usepackage{comment}
\usepackage{qcircuit}

\usepackage{tikz}
\usepackage{graphicx}
\usetikzlibrary{positioning}

\newcommand{\be}{\begin{eqnarray}}
\newcommand{\ee}{\end{eqnarray}}

\newcommand\nn{\nonumber}

\newcommand{\mat}{\left ( \begin{array}{cc}}
\newcommand{\emat}{\end{array} \right )}
\newcommand{\vect}{\left ( \begin{array}{c}}
\newcommand{\evect}{\end{array} \right )}

\definecolor{red}{rgb}{1.00, 0.00, 0.00}

\definecolor{blue}{rgb}{0.00, 0.00, 1.00}
\definecolor{green}{rgb}{0.20, 0.6, 0.1}

\definecolor{darkgreen}{rgb}{0.0, 0.4, 0.0}

\long \def \blockcomment #1\endcomment{}

\begin{document}

\title{Quantum Determinant Estimation}

\author{J.~Agerskov and K.~Splittorff}
\affiliation{\hspace{1mm} NNF Quantum Computing Programme, Niels Bohr Institute, University of Copenhagen, Denmark, Blegdamsvej 17, DK-2100, Copenhagen {\O}, Denmark}

\date{\today}
\begin{abstract}
A quantum algorithm for computing the determinant of a unitary matrix $U\in U(N)$ is given. The algorithm requires no preparation of eigenstates of $U$ and estimates the phase of the determinant to $t$ binary digits accuracy with $\mathcal{O}(N\log^2 N+t^2)$ operations and $tN$ controlled applications of $U^{2^m}$  with $m=0,\ldots,t-1$. For an orthogonal matrix $O\in O(N)$ the algorithm can determine with certainty the sign of the determinant using $\mathcal{O}(N\log^2 N)$ operations and $N$ controlled applications of $O$. An extension of the algorithm to contractions is discussed. 
\end{abstract}

\maketitle


\section{Introduction}

The ability of quantum algorithms to solve certain problems significantly faster than any known classical algorithms, has sparked immense interest in developing quantum computers. The advantage of quantum algorithms is in some cases exponential \cite{NC}, and the range of applications is ever increasing \cite{QuantumAlgorithms:survey}. A prime example is the Quantum Phase Estimation (QPE) algorithm \cite{Kitaev,CEMM,AbramsLloyd} which allows to estimate an eigenvalue of a unitary matrix to exponential accuracy. This fundamental algorithm is used in applications ranging from prime factorization \cite{Shor} to solving systems of linear equations \cite{HHL} and estimation of the ground state energy of Hamiltonians \cite{NC}. 

The aim of the present paper is to introduce a quantum algorithm which can evaluate the determinant of a unitary matrix to exponential accuracy. The motivation for developing a quantum algorithm which estimates the determinant is manifold; for example {\sl 1)} the determinant enters as a tool in many computational strategies \cite{MatrixComputations} {\sl 2)} determinants occurs naturally in partition functions where fermions have been integrated out \cite{AP} {\sl 3)} the determinant is a global property of the matrix, that is a property which is not associated with a single eigenstate of the matrix. 

The Quantum Determinant Estimation (QDE) introduced here estimates the phase of the determinant of a unitary matrix $U\in U(N)$. 
The QDE algorithm relies on two key ingredients: First, if we perform a change of basis with $U$, the completely antisymmetric state is invariant up a multiplicative factor given by the determinant of $U$. Second, the ability of the standard QPE algorithm to estimate a phase to high accuracy efficiently.  The combination of these two ingredients allows the QDE algorithm to estimate the phase of the determinant to $t$ binary digits accuracy with $\mathcal{O}(N\log^2 N+t^2)$ operations and $Nt$ controlled applications of $U^{2^m}$ {\rm where} $m=0,\ldots,t-1$. 

Note that while the QPE algorithm requires the preparation of the eigenstate belonging to the eigenvalue we wish to estimate (or at least a state with a significant overlap with this eigenstate), the QDE algorithm introduced here does not require preparation of any eigenstate of $U$. Instead it requires the preparation of a completely antisymmetric state which is independent of $U$.

Just as the order finding in Shors algorithm \cite{Shor} can be viewed as a special case of QPE for a certain unitary and a clever choice of initial state \cite{NC}, the QDE algorithm can be viewed as an application of the QPE algorithm for the matrix $U^{\otimes N}$ applied to a completely antisymmetric state. In this formulation of the QDE algorithm, we use the fact that any completely antisymmetric state, made up of a basis of the Hilbert space, is an eigenstate of $U^{\otimes N}$ with eigenvalue equal to $\det(U)$.

As an application of the QDE algorithm we show that it can determine with certainty the sign of the determinant of an orthogonal matrix with $\mathcal{O}(N\log^2 N)$ operations and $N$ controlled applications of $O$. 

A quantum algorithm for determinant estimation has been studied recently in \cite{Qi,Zenchuk} and we compare the performance of the QDE algorithm with that of \cite{Qi,Zenchuk} below. Quantum algorithms to estimate the log-determinant by means of spectral sampling have been introduced in \cite{Zhao,Luongo,Giovannetti}. We comment on assumptions and runtimes of these algorithms compared to QDE.  We introduce one possible extension of the QDE algorithm from unitary matrices to contractions, and discuss its performance. 

The paper is organized as follows: First the QDE algorithm is presented in Section \ref{sec:QDE} and in Section \ref{sec:performance} the performance of the QDE algorithm is considered. In Section \ref{sec:QDEasQPE} the QDE algorithm is reformulated as a special case of the QPE algorithm. The application of the QDE algorithm to orthogonal matrices is presented in section \ref{sec:detO}. We compare the QDE algorithm to existing algorithms in Section \ref{sec:Compare} and discuss one possible extension to non-unitary matrices in Section \ref{sec:NON-UNITARY}. Finally Section \ref{sec:Summary} contains a summery and outlook.

\section{Quantum Determinant Estimation}
\label{sec:QDE}

\noindent
{\bf The task:} Given a unitary matrix $U\in U(N)$, the task is to provide an estimate of the determinant  
\begin{eqnarray}
\label{detU}
\det(U) = e^{i\phi_U} \ .
\end{eqnarray}
The QDE algorithm will provide an estimate of $\phi_U$ which is accurate to $t$-binary digits.

\begin{figure}[ht]
	\centerline{
\Qcircuit @C=1em @R=.7em {
 \lstick{\ket{0}}   & \multigate{4}{\rm QFT} & \qw & \ctrl{+6}   & \qw         & \qw         & \qw         &\multigate{4}{{\rm QFT}^{-1}} & \qw & \meter \\
 \lstick{\ket{0}}   & \ghost{\rm QFT}           & \qw & \qw \qwx &  \ctrl{+5}  & \qw         & \qw         &\ghost{\rm QFT^{-1}}           & \qw & \meter  \\
 \lstick{\vdots \ }   &           &  &  &      &  &    \lstick{\ddots \ \ \ \ \ \ \ }    &        & &   \lstick{\; \vdots } \\
   &           &  &  &   &   &       &          & &  \\
 \lstick{\ket{0}}   & \ghost{\rm QFT}           & \qw &\qw \qwx  &  \qw \qwx & \qw         & \ctrl{+2} & \ghost{\rm QFT^{-1}}           & \qw & \meter \gategroup{2}{1}{3}{1}{5.6em}{\{}\inputgrouph{2}{3}{1em}{\text{Reg.~1}}{5.6em}\\
                         &                                &        & \qwx       &    \qwx      &                &               &                                     &        & \\
 \lstick{\ket{1}}  &  \multigate{5}{\rm Asym}                       & \qw & \multigate{5}{(U^{2^0})^{\otimes N}} & \multigate{5}{(U^{2^1})^{\otimes N}} & \qw & \multigate{5}{(U^{2^{t-1})^{\otimes N}}} &\qw   &\qw   & &  \\ 
 \lstick{\ket{2}}    &  \ghost{\rm Asym}                       & \qw & \ghost{(U^{2^0})^{\otimes N}} & \ghost{(U^{2^1})^{\otimes N}} & \qw & \ghost{(U^{2^{t-1})^{\otimes N}}} &\qw  &\qw   & &  \\
 \lstick{\vdots \ }    &           &  &  &   &   &       &          & &  \\
   \\
   &           &  &  &   &   &       &          & &  \\
 \lstick{\ket{N}}   &  \ghost{\rm Asym}                       & \qw & \ghost{(U^{2^0})^{\otimes N}} & \ghost{(U^{2^1})^{\otimes N}} & \qw & \ghost{(U^{2^{t-1})^{\otimes N}}} &\qw  &\qw   & &  \  \gategroup{8}{1}{10}{1}{5.6em}{\{}\inputgrouph{8}{3}{1em}{\text{Reg.~2}}{5.6em}
 }
}
\caption{\label{Fig:QDE} Quantum circuit description of the quantum determinant estimation (QDE) algorithm, which estimates the phase of the determinant of a unitary matrix $U\in U(N)$. The $N\log(N)$ qubits of the second register are first put into a completely antisymmetric state. Acting upon this state with $U^{\otimes N}$ results in a factor $\det(U)=e^{i\phi_U}$ according to the identity (\ref{detU-identity}). Finally the QPE algorithm is used to evaluate $\phi_U$. }
\end{figure}
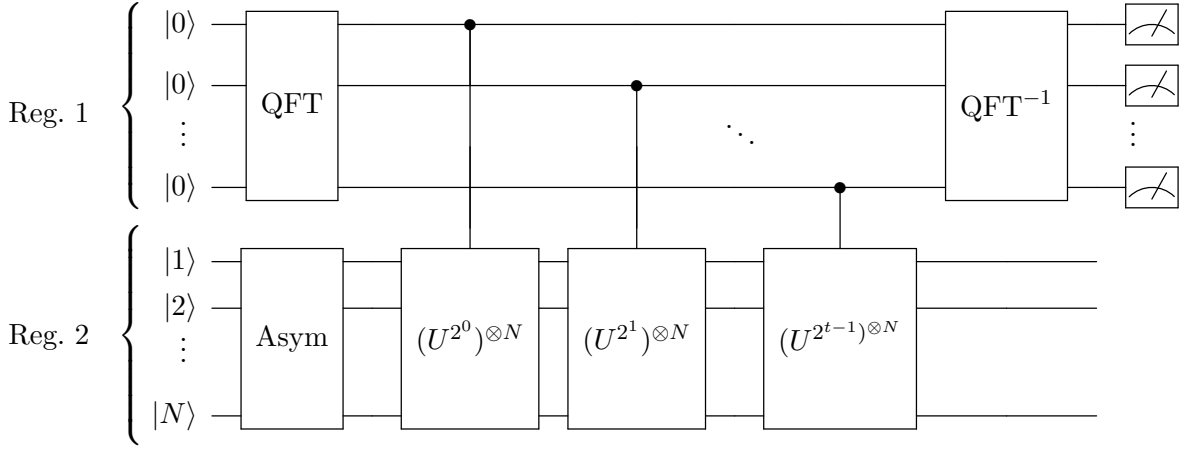

\noindent
{\bf The algorithm:} The key ingredient in the QDE algorithm is the identity
\begin{eqnarray}
\label{detU-identity}
\sum_{\sigma\in S_N} {\rm sgn}(\sigma)|U\sigma(1),\ldots,U\sigma(N)\rangle= \det(U) \sum_{\sigma\in S_N} {\rm sgn}(\sigma)|\sigma(1),\ldots,\sigma(N)\rangle \ .
\end{eqnarray}
Here $S_N$ denotes the symmetric group over the set $\{1,\ldots,N\}$ and ${\rm sgn}(\sigma)$ is the sign of the element $\sigma\in S_N$. The identity states that a completely antisymmetric tensorspaceproduct of any basis of the Hilbert space, transforms trivially under $U$ up to multiplication by the determinant of $U$.  As the determinant of a unitary matrix is a complex phase, $\det(U) = e^{i\phi_U}$, we can combine the identity (\ref{detU-identity}) with a slightly modified form of the QPE algorithm and get an estimate of $\phi_U$ to $t$ binary digits accuracy. The steps of the QDE algorithm are given in Table \ref{Tab:QDE} and the corresponding quantum circuit is displayed in Figure \ref{Fig:QDE}. The algorithm employs two registers: Register 1 with $t$ qubits and Register 2 with $N\log(N)$ qubits. 
\bigskip

The identity (\ref{detU-identity}) is sometimes taken to be the very definition of the determinant, see eg.~\cite{Bourbaki}, and it holds for matrices beyond unitary. For completeness we demonstrate in Appendix \ref{App:det-identity} that (\ref{detU-identity}) is consistent with a perhaps more familiar expression for the determinant.   
\bigskip

\begin{table}[h]
\begin{center}
\begin{tabular}{ l l l l}
{\bf Operation} & {\bf State} & {\bf Operations} & {\bf Reference} \\
\hline
 Initial & $|0\rangle|0,\ldots,0\rangle$ &  & \\ 
 Orh.n. & $|0\rangle|1,\ldots,N\rangle$  & $\mathcal{O}(N\log (N/e))$ & \\ 
 QFT & $\frac{1}{\sqrt{2^t}}\sum_{j=0}^{2^t-1}|j\rangle|1,\ldots,N\rangle$ & $\mathcal{O}(t)$  &  \cite{NC} \\  
 Asym & $\frac{1}{\sqrt{2^t}}\sum_{j=0}^{2^t-1}|j\rangle \frac{1}{\sqrt{N!}}\sum_\sigma {\rm sgn}(\sigma)|\sigma(1),\ldots,\sigma(N)\rangle$ & $\mathcal{O}(N\log^2 N)$ & \cite{Berry}  \\  
 
 $cU^{\otimes N}$  &  $\frac{1}{\sqrt{2^t}}\sum_{j=0}^{2^t-1}|j\rangle| \frac{1}{\sqrt{N!}}\sum_\sigma {\rm sgn}(\sigma)|U^j\sigma(1),\ldots,U^j\sigma(N)\rangle$  & $\mathcal{O}(tN)$  $cU^{2^m}$ & \\
  &  $=\frac{1}{\sqrt{2^t}}\sum_{j=0}^{2^t-1}e^{i\phi_U j}|j\rangle| \frac{1}{\sqrt{N!}}\sum_\sigma {\rm sgn}(\sigma)|\sigma(1),\ldots,\sigma(N)\rangle$  \\
 QFT$^{-1}$ & $\frac{1}{2^t}\sum_{j,k=0}^{2^t-1}e^{i(\phi_U-2\pi\frac{k}{2^t}) j}|k\rangle \frac{1}{\sqrt{N!}}\sum_\sigma {\rm sgn}(\sigma)|\sigma(1),\ldots,\sigma(N)\rangle$ & $\mathcal{O}(t^2)$ & \cite{NC}\\
 Measure & $|k'\rangle \frac{1}{\sqrt{N!}}\sum_\sigma {\rm sgn}(\sigma)|\sigma(1),\ldots,\sigma(N)\rangle$ & $\mathcal{O}(t)$ & \\
 \hline
\end{tabular}
\end{center}
\caption{\label{Tab:QDE} The Quantum Determinant Estimation (QDE) algorithm gives an estimate for the phase $\phi_U$ of the determinant $\det(U)=e^{i\phi_U}$ of a unitary matrix $U\in U(N)$. The estimate is accurate to $t$ binary digits where $t$ is the number of qubits in the register 1.}
\end{table}

\noindent
{\bf Assumptions:}  As the QDE algorithm makes use of a slightly modified form of QPE we assume, as is standard for QPE \cite{NC,CEMM}, that we have available black boxes capable of performing controlled $U^{2^m}$ for $m=0,\ldots,t-1$. Note, however, that we {\sl do not} need to assume that a black box exits which can prepare eigenstates of $U$, as usually required for QPE.

\section{Performance of the QDE algorithm}
\label{sec:performance}

As we now show the QDE algorithm estimates the phase of the determinant of a unitary matrix $U\in U(N)$ to $t$ binary digits using $\mathcal{O}(N\log^2 N+t^2)$ operations and $tN$ applications of $U^{2^m}$. We consider one step of the QDE algorithm at a time:

\bigskip

\noindent
{\bf Orthonormalization Reg.~2:}  To take the 2nd register from the initial state $|0,\ldots,0\rangle$ to $|1,\ldots,N\rangle$. Assume that $N=2^n$ such that each of the $N$ states can be represented by $n=\log(N)$ qubits. To encode the $j$'th element in $|1,\ldots,N\rangle$ requires of order $\log(j)$ operations, so we need in total of order $\log(N!)$ operations. For large $N$ we can use Stirling's approximation $N!= \sqrt{2\pi N} (N/e)^N( 1 + \mathcal{O}(1/N) )$ to rewrite this as $\mathcal{O}(N\log(N/e))$.
\bigskip

\noindent
{\bf QFT Reg.~1:} As the initial state of Reg.~1 is $|0\rangle$ the QFT hereof can be carried out by $H^{\otimes t}$, ie.~$t$ operations.
\bigskip

\noindent
{\bf Antisymmetrization Reg.~2:} With the algorithm of \cite{Berry} the transformation
\begin{equation}
|1,\ldots,N\rangle  \rightarrow  \frac{1}{\sqrt{N!}}\sum_\sigma {\rm sgn}(\sigma)|\sigma(1),\ldots,\sigma(N)\rangle
\end{equation}
can be carried out with a gate count of $\mathcal{O}(N\log^2 N)$. 
\bigskip
\bigskip

\noindent
{\bf Controlled unitarys Reg.~1+2:} The application of the controlled unitary operations follow that of the ordinary QPE algorithm \cite{CEMM}, only each time a $U^{2^m}$ for ($m=0,\ldots, t-1$) is applied in the QPE the exact same matrix is applied $N$ times in the QDE algorithm. This step therefore requires $N$ times as many controlled applications in the QDE algorithm as it does in the QPE algorithm, that is $tN$ controlled $U^{2^m}$ operations.  Just as for the QPE algorithm \cite{NC} the effectivity of the QDE algorithm relies on an efficient procedure to implement the controlled $U^{2^m}$ operations.
\bigskip

\noindent
{\bf QFT$^{-1}$ Reg.~1:} The inverse QFT of register one must be of the general form which requires $\mathcal{O}(t^2)$ operations \cite{NC}.
\bigskip\\
In summary, the QDE algorithm needs $\mathcal{O}(N\log^{2}(N)+t^{2})$ operations and $tN$ controlled applications of $U^{2^m}$ to estimate the determinant to $t$ binary digits accuracy. 

\section{The QDE algorithm as a special case of the QPE algorithm}
\label{sec:QDEasQPE}

One can view the QDE algorithm as a special case of the QPE algorithm where the unitary matrix in question is $U^{\otimes N}$ with $U\in U(N)$. If we re-express the identity (\ref{detU-identity}) as
\begin{eqnarray}
\label{detU-identity-as-ev-eq}
U^{\otimes N}\sum_{\sigma\in S_N} {\rm sgn}(\sigma)|\sigma(1),\ldots,\sigma(N)\rangle= \det(U) \sum_{\sigma\in S_N} {\rm sgn}(\sigma)|\sigma(1),\ldots,\sigma(N)\rangle \ ,
\end{eqnarray}
we see that a completely antisymmetric state is an eigenstate of any $U^{\otimes N}$. The associated eigenvalue is the determinant, $\det(U)=e^{i\phi_U}$, and the QPE algorithm efficiently estimates this to $t$ binary digits. Note that the preparation of the second register is independent of $U$.  The initial state of the second register is prepared in \emph{the} completely antisymmetric state. This state is independent of the basis it is expressed in. Hence the computational basis from which it is prepared is arbitrary.

\section{Determining the sign of $\det(O)$ with $O\in O(N)$.}
\label{sec:detO}

As an application we here show that the QDE algorithm can efficiently determine the sign of the determinant of an orthogonal matrix. 
The determinant of an orthogonal matrix $O\in O(N)$ is either 1 or -1, and the sign determines the class of orthogonal matrices to which $O$ belongs: If the determinant is 1 then $O$ can be continuously deformed to the identity, however, if the determinant is -1 a reflection is needed before the matrix can be deformed continuously to the identity. 

Since the orthogonal group $O(N)$ is a subgroup of the unitary group $U(N)$ we can use the QDE algorithm to estimate the sign of the determinant. In fact since $\det(O)$ is known to be either 1 or -1 we can determine this sign with certainty using the QDE algorithm with $t=1$. To see this we write out the steps and for brevity introduce the shorthand 
\be
\label{ASYM}
\ket{\rm ASYM}= \frac{1}{\sqrt{N!}}\sum_{\sigma\in S_n} {\rm sgn}(\sigma)|\sigma(1),\ldots,\sigma(N)\rangle \ .
\ee

\bigskip

\begin{tabular}{ l l }
{\bf Operation} & {\bf State} \\
\hline
 Initial & $|0\rangle|0,\ldots,0\rangle$  \\ 
 Orh.n. & $|0\rangle|1,\ldots,N\rangle$   \\ 
 $H\otimes$ Asym & $\frac{1}{\sqrt{2}}(|0\rangle+|1\rangle)|{\rm ASYM}\rangle$  \\  
 $cO^{\otimes N}$ &  $\frac{1}{\sqrt{2}}(|0\rangle|{\rm ASYM}\rangle + |1\rangle O^{\otimes N}|{\rm ASYM}\rangle)$   \\
  &  $=\frac{1}{\sqrt{2}}(|0\rangle+\det(O)|1\rangle) |{\rm ASYM}\rangle$  \\ 
 H & $(\frac{1}{2}(1+\det(O))|0\rangle+\frac{1}{2}(1-\det(O))|1\rangle)|{\rm ASYM}\rangle$ \\ 
 \hline
\end{tabular}
\bigskip

\begin{figure}[ht]
	\centerline{
\Qcircuit @C=1em @R=.7em 
{
 \lstick{\ket{0}}                 & \gate{H} & \ctrl{+1}                      & \qw & \gate{H}  & \qw & \meter \\ 
  \lstick{\ket{\rm ASYM}}  & \qw        & \gate{O^{\otimes N}} & \qw & \qw          & \qw  &  &  &   \lstick{\ket{\rm ASYM}}   \\ 
 }
}
\caption{\label{Fig:QDE-for-O} Application of the QDE algorithm to orthogonal matrices $O\in O(N)$. The quantum circuit determines with certainty if $\det(O)=1$ or $-1$. }
\end{figure}
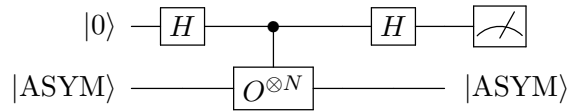

\noindent
If the final measurement on the single qubit in the first register is 0 then $\det(O)=1$ and if the result of the measurement is 1 then $\det(O)=-1$. Hence the QDE algorithm with certainty determines the sign of $\det(O)$ with $\mathcal{O}(N\log^2 N)$ operations and $N$ controlled applications of $O$. A quantum circuit describing the algorithm which determines $\det(O)$ is given in figure \ref{Fig:QDE-for-O}. 
\bigskip

\section{Comparison of QDE to existing quantum algorithms}
\label{sec:Compare}

Let us now compare the QDE algorithm to existing quantum algorithms for the computation of the determinant. 

\subsection{The QDE algorithm vs.~the QPE algorithm for each eigenvalue}
 
One could attempt to compute the determinant of $U$ by applying the QPE algorithm to estimate each of the $N$ eigenvalues of $U$ and then take their product to obtain the determinant. This can be done with $Nt$ applications of $U^{2^m}$, $m\in\{1,\ldots ,t-1\}$, and of order $Nt^2$ other operations. This procedure, however, requires that we can accurately prepare each of the $N$ eigenstates of $U$, a possibly challenging task for large $N$. 
In comparison, the QDE algorithm, requires us to prepare any $N$ orthonormal states which spans the Hilbert space, a task with no reference to the eigenvectors of $U$.

\subsection{Other algorithms for estimating the determinant}

Recently a quantum algorithm for calculating the determinant of an $N\times N$ matrix was studied in \cite{Qi}. Their technique involves encoding the columns of a matrix into quantum states. These ideas were further developed in \cite{Zenchuk}. The algorithms in \cite{Qi, Zenchuk}, are valid for a broader class of matrices than unitary matrices, but requires normalization conditions on the columns to be satisfied. The reported depth of the algorithm of \cite{Zenchuk} is $\mathcal{O}(N\log^2N)$ or $\mathcal{O}(N^2\log N)$ in the worst case, however the algorithm depends on a measurement on ancilla qubits. The probability of obtaining the correct measurement result in this step decrease exponentially with $N$. Therefore this algorithm will in practice need to be iterated an exponential number of times to succeed \cite{Zenchuk}.

The QDE algorithm only requires a single final measurement in the first register, however at present the QDE algorithm only applies to unitary matrices. One possible extension of the QDE algorithm to non-unitary matrices is discussed below in section \ref{sec:NON-UNITARY}. This particular extension depends on a measurement on ancilla qubits and the probability of obtaining the correct measurement result decrease exponentially with the precision $t$.

\subsection{Algorithms for estimating the log determinant}

Quantum algorithms for estimating the log-determinant with a spectral sampling protocol were developed in \cite{Zhao, Luongo, Giovannetti}. Here one performs QPE with a fully mixed state as input, in order to uniformly sample the log-eigenvalues until a good estimate for the mean of the log-eigenvalues is found. For these protocols one can obtain $\log(N)$ scaling in the number of operations. However, the setting is quite different from the assumptions made for the QDE algorithm. For the log-determinant to be well defined the matrix in question must be positive definite. In \cite{Giovannetti}, it is assumed that the matrix is positive definite with sparsity $s$ and with the spectrum of the matrix is contained in $(1/(2\kappa),1/2)$ for some $\kappa>1$. Under these assumptions the log-determinant can be estimated with a relative error $\epsilon$ in $\mathcal{O}\left(\frac{\kappa \log^2(\kappa)s^2}{\epsilon^3}\log(N)\log^2(s^2/\epsilon)\right)$ steps \cite{Giovannetti}. Note that a relative error of the log-determinant, $\epsilon$, gives a relative error on the determinant of order $e^{N\epsilon}$ under these assumptions, as the log-determinant itself scales with $N$.

As noted the assumptions for estimating the log-determinant are quite different from those of QDE  and this hinders a direct comparison. In order to make an indirect comparison we now consider the runtime of QDE for a unitary operator $U=\exp(iH)$ where $H$ is a sparse and Hermitian matrix. The QDE algorithm is highly parallelizable: The orthonormalization can be carried out completely in parallel with a runtime $\mathcal{O}(1)$, the first quantum Fourier transform consists of $t$ Hadamard gates which can be done in parallel ie.~again with a runtime $\mathcal{O}(1)$, the antisymmetrization is also parallelizable \cite{Berry} with a runtime  $\mathcal{O}(\log(N)\log(\log(N)))$, as explained in \cite{Giovannetti} the controlled unitary operations for the sparse matrix can be carried out withe a runtime $\mathcal{O}(\log(N)/\epsilon)$, where $\epsilon$ is the precision on phase of the determinant, and finally the inverse quantum Fourier transform can be carried out with a runtime of order $\mathcal{O}(t\log(t))$. This results in a total runtime of QDE for such a unitary matrix scaling as $\log(N)/\epsilon+\log(N)\log(\log(N))$.

\section{Generalization to non-unitary matrices}
\label{sec:NON-UNITARY}

\begin{figure}[ht]
	\centerline{
\Qcircuit @C=1em @R=.7em {
 \lstick{\ket{0}}   & \multigate{4}{\rm QFT} & \qw & \ctrl{+6}   &\qw& \qw         & \qw         & \qw         &\multigate{4}{{\rm QFT}^{-1}} & \qw & \meter \\
 \lstick{\ket{0}}   & \ghost{\rm QFT}           & \qw & \qw \qwx &\qw &\ctrl{+5}  & \qw         & \qw         &\ghost{\rm QFT^{-1}}           & \qw & \meter  \\
 \lstick{\vdots \ }   &           &  &  &  &    &   &   \lstick{\ddots \ \ \ \ \ }    &        & &   \lstick{\; \vdots } \\
   &           &  &  & &   &   &       &          & &  \\
 \lstick{\ket{0}}   & \ghost{\rm QFT}           & \qw &\qw \qwx  &\qw&  \qw \qwx & \qw         & \ctrl{+2} & \ghost{\rm QFT^{-1}}           & \qw & \meter \gategroup{2}{1}{3}{1}{5.6em}{\{}\inputgrouph{2}{3}{1em}{\text{Reg.~1}}{5.6em}\\
                         &                                &        & \qwx       &  &  \qwx      &                &               &                                     &        & \\
 \lstick{\ket{1}}  &  \multigate{5}{\rm Asym}                       & \qw & \multigate{7}{\operatorname{U}(A^{2^0})} &\qw& \multigate{9}{\operatorname{U}(A^{2^1})} & \qw & \multigate{12}{\operatorname{U}(A^{2^{t-1}})} &\qw   &\qw   & &  \\ 
 \lstick{\ket{2}}    &  \ghost{\rm Asym}                       & \qw & \ghost{\operatorname{U}(A^{2^0})} &\qw& \ghost{\operatorname{U}(A^{2^1})} & \qw & \ghost{\operatorname{U}(A^{2^{t-1}})} &\qw  &\qw   & &  \\
 \lstick{\vdots \ }    &           &  &  & &   &   &       &          & &  \\
   \\
   &           &  &  & &  &   &       &          & &  \\
 \lstick{\ket{N}}   &  \ghost{\rm Asym}                       & \qw & \ghost{\operatorname{U}(A^{2^0})} &\qw& \ghost{\operatorname{U}(A^{2^1})}  & \qw & \ghost{\operatorname{U}(A^{2^{t-1}})}&\qw  &\qw   & &  \  \gategroup{8}{1}{10}{1}{5.6em}{\{}\inputgrouph{8}{3}{1em}{\text{Reg.~2}}{5.6em}\\
                         &                                &        &       &        &                &               &                                     &        & \\
 \lstick{\ket{0}}  &  \qw                     & \qw &  \ghost{\operatorname{U}(A^{2^0})} & \measureD{0}   \\ 
 \lstick{\ket{0}}    &  \qw                  & \qw & \qw & \qw& \ghost{\operatorname{U}(A^{2^1})}  &\measureD{0}   \\
 \lstick{\vdots \ }    &           &  &  &   &   &       &          & &  \\
   \\
   &           &  &  &   &   &       &          & &  \\
 \lstick{\ket{0}}   &  \qw                      & \qw & \qw & \qw& \qw & \qw&\ghost{\operatorname{U}(A^{2^{t-1}})} &\measureD{0}  \  \gategroup{15}{1}{17}{1}{5.6em}{\{}\inputgrouph{15}{3}{1em}{\text{Reg.~3}}{5.6em}
 }
}
\caption{\label{Fig:QDE3} Quantum circuit description of one possible extension of the quantum determinant estimation (QDE) algorithm to contractions $A\in GL(N)$. As indicated the algorithm requires that the result of the measurement on the ancillas is zero across the third register.}
\end{figure}
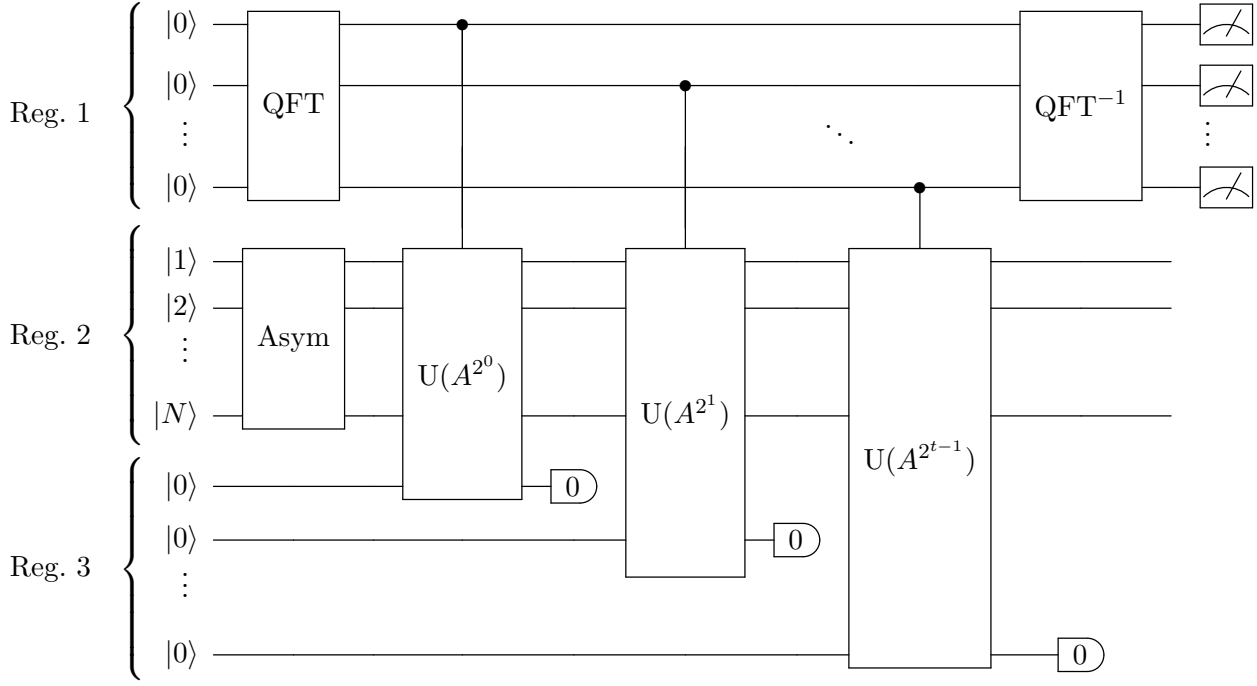

Here we discuss one possible way to generalize the QDE algorithm to a broader class of matrices, namely contractions, that is matrices $A$ with norm $||A||\leq1$. 

To extend the QDE algorithm to contractions we first note that contractions can be block-encoded in a unitary matrix
\begin{eqnarray}
U(A)=\left(\begin{array}{cc} A & (1-AA^\dagger)^{1/2} \\ (1-A^\dagger A)^{1/2} & -A^\dagger \end{array}\right) \ .
\end{eqnarray}
If $A$ is of size $N\times N$ we can therefore construct an encoding, $U$, using one additional ancilla qubit, such that $A=\langle 0 | U | 0\rangle$. Using this and the shorthand $|{\rm ASYM}\rangle$ introduced in (\ref{ASYM}) we have 
\begin{eqnarray}
\label{U(A)action}
U(A)|{\rm ASYM}\rangle\otimes|0\rangle & = & A|{\rm ASYM}\rangle\otimes|0\rangle + (1-A^\dagger A)^{1/2}|{\rm ASYM}\rangle\otimes|1\rangle \\ 
  & = & \det(A)|{\rm ASYM}\rangle\otimes|0\rangle + \det(1-A^\dagger A)^{1/2}|{\rm ASYM}\rangle\otimes|1\rangle \ . \nn
\end{eqnarray}
As $U(A)$ is unitary the state remains normalized and hence the probability to measure zero on the ancilla is $P_0=|\det(A)|^2$. If the measurement is 0 the state after the measurement is $(\det(A)/|\det(A)|) |{\rm ASYM}\rangle\otimes|0\rangle$. Note that $\det(A)/|\det(A)|$ is the phase $e^{i\phi}$ of the determinant of $A$, $\det(A)=re^{i\phi}$, and that we can use the QPE algorithm to estimate this phase. The QDE algorithm can therefore be extended to contractions as described in the quantum circuit of figure \ref{Fig:QDE3}.
The measurement of the first register of the circuit in Figure \ref{Fig:QDE3} estimates the phase of the of determinant of $A$, and the estimate will be precise to $t$ digits. However, as indicated in the quantum circuit the algorithm is only successful provided that the measurement on the ancillas in register 3 is 0. The probability of measuring 0 all across register 3 is
\begin{eqnarray}
P_{0,\ldots,0} & = & P_0(A^{2^0})\cdot\ldots\cdot P_0(A^{2^{t-1}}) \\ 
                       & = & |\det(A^{2^0})|^2\cdot\ldots\cdot |\det(A^{2^{t-1}})|^2\nn \\
                       & = & |\det(A)|^{2(2^t-1)} \ . \nn
\end{eqnarray}
As soon as $|\det(A)|<1$ the probability $P_{0,\ldots,0}$ becomes very small in $t$, hence the quantum circuit must be repeated a large number of times. This particular extension of the QDE algorithm from unitary matrices to contractions hence comes with an exponential overhead. 
However, given the measurement statistics from the repeated algorithm, the magnitude of the determinant can be estimated. 

\section{Summary and outlook}
\label{sec:Summary}

We have introduced a quantum algorithm which estimates the determinant of a unitary matrix $U\in U(N)$. The algorithm makes use of the fact that under a change of basis by $U$ a completely antisymmetric state transforms into itself times the determinant of $U$. For unitary matrices the determinant is a phase and hence a slightly modified version of the quantum phase estimation algorithm can be applied to accurately estimate this phase with high efficiency. The QDE algorithm can also be seen as a special case of the QPE algorithm for the matrix $U^{\otimes N}$. Note that no preparation of eigenstates of $U$ is required for the QDE algorithm.

From the perspective of classical algorithms the direct application of the central identity (\ref{detU-identity}) in the QDE algorithm appears not to be effective.  However, the QDE algorithm inherits the speedup of the QPE algorithm, leading to an estimate of $\phi_U$ which is correct to $t$-binary digits with $\mathcal{O}(N\log^2 N+t^2)$ operations and $tN$ applications of $U^{2^m}$, $m=0,\ldots,t-1$. As for any application of QPE \cite{NC} the efficiency of the algorithm depends on the ability to apply the controlled $U$ operations.

We have applied the QDE algorithm to orthogonal matrices, and shown that it can determine with certainty the sign of the determinant using $\mathcal{O}(N\log^2 N)$ operations and $N$ controlled applications of $O$. The QDE algorithm may also find applications for computations of partition functions with a fermion determinant, in particular in the presence of a sign problem where accurate estimates of the determinant are essential \cite{sign-problem}.

A central part of the QDE algorithm is the anti-symmetrization of the initial state of the second register. Antisymmetric states are essential for many applications to chemistry and it would be most interesting to study the interplay of the QDE algorithm and the preparation of completely antisymmetric states in further detail. In addition it would be interesting to examine if there exists a simple physical system which realize QDE, as has been found for QPE in \cite{QPE-AB}. 

Finally we suggested one possible extension of the QDE algorithm to contractions. This particular generalization has an exponential overhead in $t$ as soon as the determinant is not of unit magnitude. An interesting open problem, is to examine if a generalization that scales better for non-unitary matrices exists. The result of \cite{Dorn} suggests that such an extension cannot make due with less that $N^{2}$ queries to the matrix, if the algorithm can determine whether or not the determinant is zero.

\bigskip

\acknowledgments
This work is supported by the Novo Nordisk Foundation, Grant number NNF22SA0081175, NNF Quantum Computing Programme. 

\newpage 

\appendix
\section{Identity (\ref{detU-identity})}
\label{App:det-identity}

The central identity \eqref{detU-identity} is a special case of the general identity 
\begin{eqnarray}
\label{detA-identity}
\sum_{\sigma\in S_N} {\rm sgn}(\sigma)|A\sigma(1),\ldots,A\sigma(N)\rangle= \det(A) \sum_{\sigma\in S_N} {\rm sgn}(\sigma)|\sigma(1),\ldots,\sigma(N)\rangle \ ,
\end{eqnarray}
which holds for any $N\times N$ matrix. This identity is sometimes taken to be the very definition of the determinant \cite{Bourbaki}. Here we show that \eqref{detA-identity} is consistent with the perhaps more familiar relation 
\begin{equation}\label{Eq:Determinant1}
\sum_{i_1,i_2,...,i_N}\epsilon_{i_1 i_2\ldots i_N}A_{1 i_1} A_{2 i_2}\ldots A_{N i_N}=\det(A),
\end{equation}
where $A_{ji}=\braket{j\vert A \vert i}$ and $\epsilon$ is the Levi-Civita symbol. First, observe that \eqref{Eq:Determinant1} is antisymmetric under exchange of the row-indices of the matrices, hence it follows from \eqref{Eq:Determinant1} that
\begin{equation}\label{Eq:Determinant2}
\sum_{i_1,i_2,...,i_N}\epsilon_{i_1 i_2\ldots i_N}A_{j_1 i_1} A_{j_2 i_2}\ldots A_{j_N i_N}=\det(A)\epsilon_{j_1 j_2\ldots j_N} \ .
\end{equation}
Next, the expression \eqref{Eq:Determinant2} is equivalent to \eqref{detA-identity} which follows if we take the inner product of \eqref{detA-identity} with $\bra{j_{1},j_{2},...,j_{N}}$ .


\end{document}